\documentclass[twocolumn,preprintnumbers,showpacs]{revtex4-1}
\usepackage{amsmath}
\usepackage{amssymb}
\usepackage{graphicx}
\usepackage{color}
\usepackage{float}
\usepackage{sistyle}
\usepackage{subfigure}  
\usepackage{verbatim}   
\usepackage{graphicx}
\usepackage{dcolumn}
\usepackage{bm}
\usepackage[citecolor=blue,filecolor=blue,colorlinks=true,urlcolor=blue]{hyperref}
\usepackage{ulem}

\newcommand{\BAN}{\ensuremath{B_{1g}\,}}

\newcommand{\Tc}{\ensuremath{T_{\rm c}\,}}

\newcommand{\cm}{\ensuremath{{\rm cm}^{-1}}}

\def\g{\gamma}
\def\w{\omega}

\graphicspath{{images/}{../images/}}

\begin{document}

\title{Unconventional high-energy-state contribution to the Cooper pairing in under-doped copper-oxide superconductor HgBa$_2$Ca$_2$Cu$_3$O$_{8+\delta}$}

\author{B. Loret$^1$, S. Sakai $^2$, Y. Gallais$^1$, M. Cazayous$^1$, M.-A. M\'easson$^1$, A. Forget$^4$, D. Colson$^4$, M. Civelli$^3$ and A. Sacuto$^1$ }

\affiliation{$^1$ Laboratoire Mat\'eriaux et Ph\'enom$\grave{e}$nes Quantiques (UMR 7162 CNRS), Universit\'e Paris Diderot-Paris 7, Bat. Condorcet, 75205 Paris Cedex 13, France\\
$^2$ Center for Emergent Matter Science, RIKEN, 2-1 Hirosawa, Wako, Saitama 351-0198, Japan\\
$^3$ Laboratoire de Physique des Solides, CNRS, Univ. Paris-Sud, Universit\'e Paris-Saclay, 91405 Orsay Cedex, France\\
$^4$ Service de Physique de l'{\'E}tat Condens{\'e}, DSM/IRAMIS/SPEC (UMR 3680 CNRS), CEA Saclay 91191 Gif sur Yvette cedex France}

\date{\today}


\begin{abstract}

We study the temperature-dependent electronic \BAN Raman response of a 
slightly under-doped single crystal HgBa$_2$Ca$_2$Cu$_3$O$_{8+\delta}$ with a 
superconducting critical temperature \Tc=122 K. Our main finding is that the superconducting
pair-breaking peak is associated with a dip on its higher-energy side, disappearing together
at \Tc. This result hints at an unconventional pairing mechanism, whereas spectral
weight lost in the dip is transferred to the pair-breaking peak at lower energies. This conclusion
is supported by cellular dynamical mean-field theory on the Hubbard model, 
which is able to reproduce all the main features of the \BAN Raman response and explain the 
peak-dip behavior in terms of a nontrivial relationship between the superconducting and the pseudo gaps.  
\end{abstract}

                                                     
\maketitle

Conventional superconductors are well understood within the Bardeen-Cooper-Schrieffer (BCS) theory \cite{Bardeen57}: 
below a critical transition temperature \Tc, electrons at a characteristic energy (the Fermi energy) 
bind into Cooper pairs by an effective attractive interaction mediated by lattice 
vibrations (phonons) \cite{Schrieffer1969}. The Bose condensate of pairs displays then zero resistance to electrical conduction and a gap opens in spectroscopic observables by a transfer of spectral weight from the Fermi level to higher energies.
The BCS pairing mechanism, however, has not been able to account for 
the high \Tc observed in copper-oxide (cuprate) superconductors.
In these materials the isotopic effect is extremely weak and does not suggest a strongly 
coupled phonon-mediated superconductivity \cite{Batlogg1987}. 

The nature of the pairing interaction has therefore remained controversial. Possible proposals include strong electronic correlations stemming from Mott physics \cite{Anderson87} 
or the competition with other exotic phases such as charge \cite{Kivelson2003,Chubukov2014,Meier2014}, 
spin density \cite{Demler2001,Moon2009,Scalapino2012} waves or loop currents\cite{Varma1997}. 

The scenario is further complicated by the presence of another gap (the pseudogap), which
is an ingredient missing in the BCS description. 
The pseudogap manifests itself above \Tc as a loss of quasiparticle spectral weight
\cite{Alloul1989,Warren1989,Timusk1999}. Whether or not the pseudogap plays any role in the 
high-\Tc mechanism, this remains a fundamental open question.  
This inherent complexity of the cuprates has hidden key features of the pairing mechanism in most 
experiments, preventing a satisfactory understanding of high \Tc superconductivity. 

In this article we present an electronic Raman scattering study in the \BAN geometry
on a slightly-underdoped (UD) three-copper-oxide-layer compound HgBa$_2$Ca$_2$Cu$_3$O$_{8+\delta}$ (Hg-1223). 
We reveal a nontrivial relationship between the pair-breaking peak (PP), which corresponds to two Bogoliubov quasiparticle excitations,
and a loss of spectral weight (dip) appearing 
on its higher-energy side. Remarkably, the PP and dip disappear simultaneously at \Tc, indicating a transfer of spectral weight from the dip electronic states to the PP at lower energies. 
This behavior is in sharp contrast with the BCS pairing mechanism, which involves only the low-energy electronic states around the Fermi level, being transferred to the superconducting gap edges below \Tc. We are able to explain our experimental observations using the cellular dynamical mean-field theory \cite{Kotliar2001} (CDMFT) applied to the two-dimensional Hubbard model, the basic strongly correlated electron model describing copper-oxygen planes in cuprates. 
CDMFT unveils an unusual relationship between a particle-hole symmetric superconducting gap and a particle-hole asymmetric pseudogap, coexisting below \Tc. While below the Fermi level they share the same gap edge, 
above the Fermi level they compete for the same states. Spectral weight is in fact removed from the pseudogap upper edge to lower 
energies contributing to the formation of the upper superconducting Bogoulibov peak. 
This unconventional mechanism is ultimately responsible for the PP-dip behaviour observed in the \BAN Raman response.

\begin{figure}[tb!]
\begin{center}
\includegraphics[width=9cm]{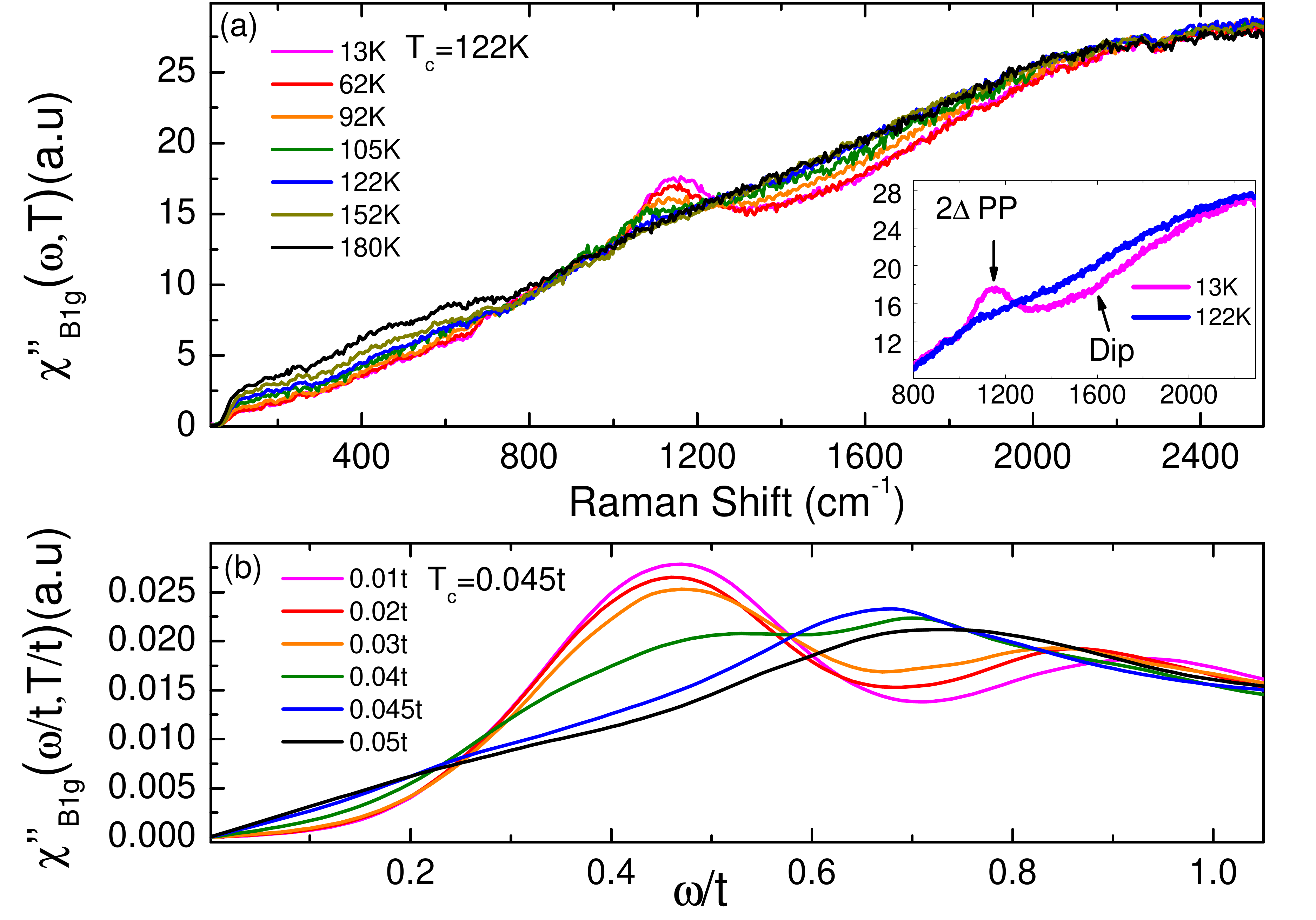}
\caption{(Color online). (a) Temperature dependence of the \BAN Raman response 
for the UD Hg-1223 single crystal (\Tc=122 K). The $2\Delta$ pair breaking peak (PP) is detected at $\approx 1135$ ~\cm and the dip at $\approx 1600$ ~\cm. In the inset, a closer view of the peak-dip structure is shown. The pink (light grey) curve was measured at 13 K and the blue (dark grey) one at 122 K. (b) Temperature dependence of the  theoretical \BAN Raman response within CDMFT (\Tc $\approx 0.045 t$). Pair breaking peak is observed at $\omega\approx 0.45t$ and the dip at $\omega\approx 0.7t$}.
\label{fig:1}
\end{center}\vspace{-7mm}
\end{figure}

Raman measurements have been carried out on UD Hg-$1223$ single crystal with \Tc=122 K grown by a single step synthesis under normal pressure in evacuated quartz tube \cite{Colson1994, Bertinotti1995}. The Hg-1223 cuprate family exhibits the highest critical temperature \Tc$=$ 135~K at ambient pressure \cite{Schilling1993}. In this material the phonons do not mask the low-energy electronic spectrum \cite{Sacuto1996,Sacuto1997,Sacuto1998,Sacuto2000a} contrary to other cuprates. This gives us an unique opportunity to resolve detailed features of the superconducting state. 
Moreover the large \Tc is suitable for studying the superconducting features over a wide temperature range. 

The \BAN-symmetry Raman response, obtained from crossed light polarizations along the Cu-O bond directions, gives us access to the antinodal region of the momentum space where the superconducting gap is maximal and the 
pseudogap sets in. All the spectra have been corrected for the Bose factor and the instrumental spectral response. They are thus proportional to the imaginary part of the Raman response function $\chi^{\prime \prime}(\omega,T)$ \cite{Sacuto2013}.  

In Fig.~\ref{fig:1} (a) the \BAN Raman response $\chi^{\prime \prime}_{\BAN} (\omega, T)$ is displayed over a wide frequency range (up to 2500 ${\rm cm}^{-1}$) and from $T$= 13 K to $T$= 180 K. The key feature that we observe in the superconducting state ($T<$ 122 K) is the PP at twice the superconducting gap $2\Delta=1135\pm10~\cm$ followed at higher Raman shift by the
dip in the electronic continuum at $1600\pm40~\cm$ ( $\omega_{dip}/2\Delta =1.4\pm0.1$). 
The PP-dip structure has also been found in a two-layer compound \cite{Chubukov1999} and 
we have checked that it also exists in the superconducting Raman response of 
a single layer material HgBa$_2$CuO$_{4+\delta}$ (Hg-1201) (see Supplemental Material). 
Therefore the PP-dip structure (that we observe in the superconducting state) is different from a bilayer band splitting effect proposed to explain the peak-dip-hump structure in the ARPES spectra of a two-layer Bi-2212  compound \cite{Feng2001,Damascelli2003}. This is confirmed by the fact that the PP-dip feature in the Raman spectrum disappears at \Tc while the band splitting effect is supposed to persist above \Tc \cite{Feng2001,Damascelli2003}. 
 
The PP position $2\Delta\approx 14k_B$\Tc is in good agreement with earlier tunneling and optical measurements on Hg-1223 compounds \cite{Wei1998, McGuire2000}. In tunneling, the gap for an optimally doped Hg-1223 was estimated to be 
$\sim13k_{B}$\Tc. In optical conductivity, the pairing gap deduced 
from the scattering rate was estimated to be about 1100 \cm  $\sim13~ k_{B}$\Tc.

\begin{figure}[tbp!]
\begin{center}
\includegraphics[width=8.5cm]{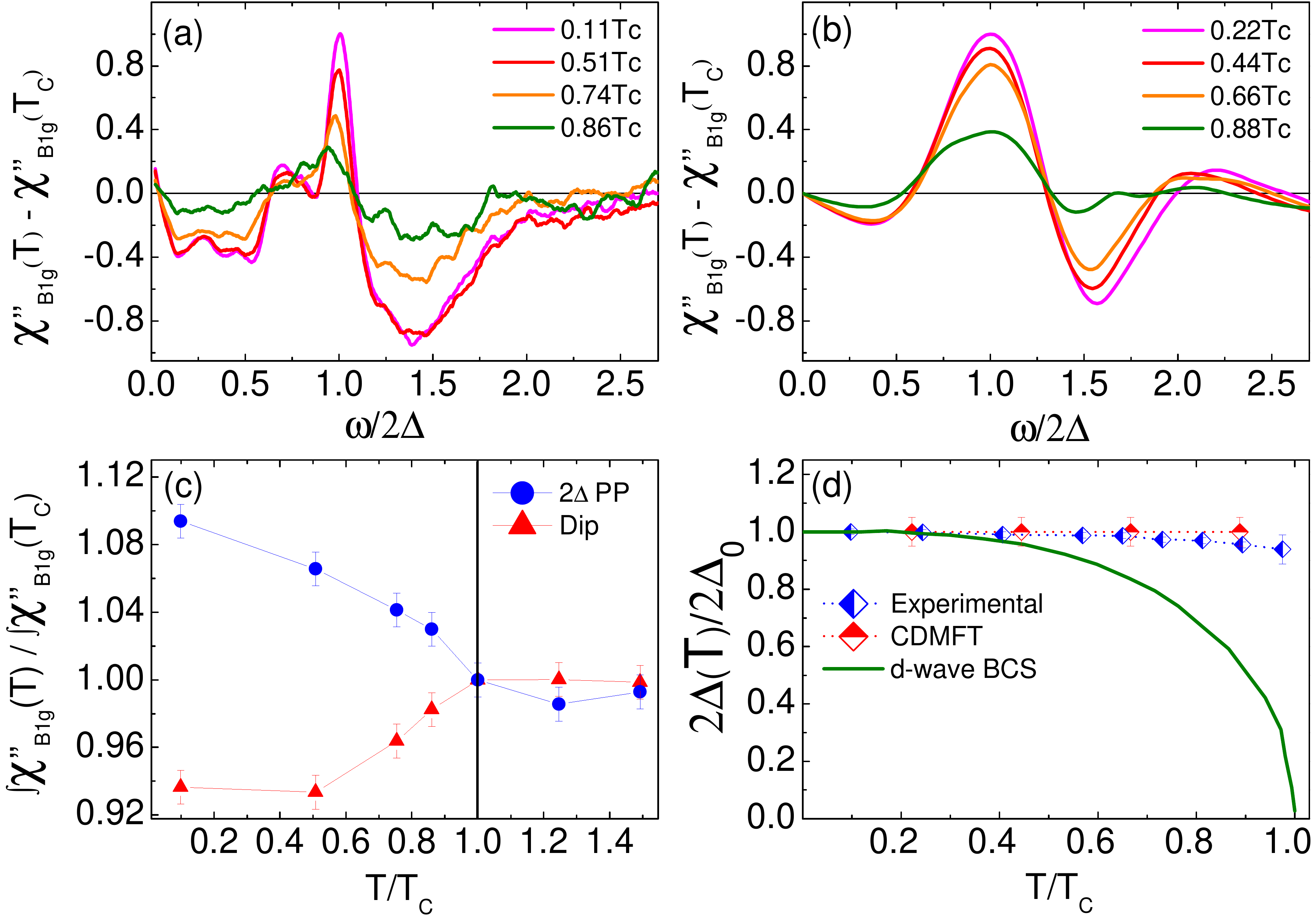}
\caption{(Color online). (a) and (b) Temperature dependence of the \BAN Raman responses subtracted from the one at \Tc for experimental data and CDMFT one. Both are normalized in intensity (with respect to the lowest temperature in the superconducting state) and in energy by $2\Delta$.  
(c) Normalized integrated intensities of the PP and the dip (obtained from Fig.~1(a)) as a function of $T/\Tc$. The PP and dip integrations extend respectively from 1000 $\cm$ to 1250 $\cm$ and from 1250 $\cm$ to 2300 $\cm$. The error bar is about $1\%$. (d) Temperature dependence of the $2\Delta$ PP obtained from experiment and the CDMFT calculations. Experimentally the error bar stemming from the spectrometer resolution is about $1\%$ except for the 105K response where the PP is broader and the error bar is about $3\%$. Theoretically the error bar is coming from the energy grid of the calculation and it is about $5\%$. The solid curve shows the temperature dependence of a $d$-wave gap in the weak coupling theory \cite{Branch1995}}.
\label{fig:2}
\end{center}\vspace{-7mm}
\end{figure}

As the temperature rises up to \Tc, the PP decreases in intensity while 
the low energy electronic background below $\approx800$~\cm ($\approx0.7$ in 2$\Delta$ unit) increases, as clearly seen by plotting the 
difference $\chi_{\BAN}(T)-\chi_{\BAN}(\Tc)$ (Fig.~\ref{fig:2} (a)) \cite{Note}. This behavior is typically expected in the BCS theory,
where the low energy spectral weight is removed and transferred to the superconducting gap edges, producing the 
2$\Delta$ PP in the $\chi_{\BAN}(T)$.
The remarkable fact in the present result is that the dip too, around $1600~\cm$, is filled up completely together with the 
decrease of the PP and disappears at \Tc=122 K. This is better seen in Fig.~\ref{fig:2} (c), where we plot the 
normalized integrated Raman intensity associated with the PP and the dip. The PP and dip lines join together 
at \Tc by definition, but the fact that they are essentially constant above \Tc shows that the two features have 
disappeared in the normal state.

Since the Raman response in Fig.~\ref{fig:1}(a) is $T$-independent above the dip energy in the superconducting state (this has been checked up to 4600 ~$\cm$), it is natural to infer that the lost spectral weight in the dip is transferred to the 2$\Delta$ PP, producing an
unconventional pairing mechanism. The possibility of high energy-state contribution to the pairing was suggested in earlier optical measurements \cite{Basov1999,Santander2003} and ARPES results\cite{Hashimoto2015}.

Another known non-BCS behavior underlined by our experimental findings is the energy location of the 
2$\Delta$ peak, which is roughly constant with increasing temperature up to \Tc (see Fig.~\ref{fig:2}(d)). 
This property is general among one and two-copper-oxide-layer compounds slightly underdoped such as Hg-1201 \cite{Guyard2008} and 
Bi$_2$Sr$_2$CaCu$_2$O$_{8+\delta}$ (Bi-2212) \cite{Staufer92}, and it is another sign of unconventional behavior.

Notice that above \Tc=122 K, the low energy background level continues to rise with $T$ (see Fig.~1 (a)). This is the Raman signature of the presence of the pseudogap in the normal state which manifests itself as a recovery of low energy spectral weight \cite{Sakai2013,Benhabib15}. 

All the above temperature-dependent features, in particular the 2$\Delta$ PP-dip structure, 
of the \BAN response are difficult to explain in the framework of BCS-like theory. 
For instance the weak-coupling BCS theory \cite{Branch1995} of a $d$-wave superconductor does explain a 2$\Delta$ peak 
which collapses at \Tc (Figs.~\ref{fig:1}(a) and \ref{fig:2}(a)),
however this should be accompanied by a consequent reduction of the peak position 2$\Delta$ 
with increasing $T$, contrary to the above experimental observation (see Fig.~\ref{fig:2} (d)). 
Moreover no dip is explicable within the BCS theory. 

The fact that 2$\Delta$ is much higher than 4.28$k_B$\Tc expected in the weak coupling theory \cite{Musaelian1996} hints at a strong coupling nature for the pairing state. This is the case for instance in preformed pair theories \cite{Anderson87,Kotliar1988,Emery1995,Franz1998,Altman2002}. In such theories incoherent Cooper pairs  exist above \Tc and their pairing gap is identified with the beforehand mentioned pseudogap.
Below \Tc the pairs acquire phase coherence establishing a superconducting state, and the pseudogap
smoothly evolves into the superconducting gap.
A main consequence is that the spectroscopic gap amplitude $\Delta$, and hence the 2$\Delta$ PP energy position, 
is only slightly temperature-dependent approaching \Tc from below \cite{Mihlin2009}.
Within such a scenario however the PP should survive above \Tc, contrary to the experimental 
observations. 
A theory describing the interaction between superconductivity and spin-density waves \cite{Chubukov1999} may account for the  
the 2$\Delta$-peak-dip feature below \Tc. Also in this case, however, the PP feature is expected to survive above \Tc.  
We show now that the CDMFT calculation of the Hubbard model can qualitatively account for all the experimental features observed above and explain the tight relationship between the PP and the dip in the Raman response.  


The Raman spectra are calculated in CDMFT within the bubble approximation through
\begin{align}
\chi_{\BAN}''(\w)=2&\int \frac{d\mathbf{k}}{(2\pi)^2} \g_{\BAN} ^2(\mathbf{k}) 
 \int_{-\infty}^\infty d\w' [f(\w')-f(\w+\w')]\nonumber\\
&\times [{\rm Im}G(\mathbf{k},\w'){\rm Im}G(\mathbf{k},\w+\w')\nonumber\\
  &-{\rm Im}F(\mathbf{k},\w'){\rm Im}F(\mathbf{k},\w+\w')]
\label{eq:raman}
\end{align}
with $\g_{\BAN}$$=$$\frac{1}{2}[\cos(k_x)-\cos(k_y)]$ and $f(\w)$ is the Fermi distribution function.
Here, the normal ($G$) and anomalous ($F$) Green's functions calculated with the CDMFT are interpolated in the momentum space \cite{Kyung2006}. This approximation is quite robust around the antinodal region, which includes the  cluster momenta $ \mathbf{K}= (0, \pm \pi)$, $(\pm \pi, 0) $, and will not affect our conclusions on the \BAN Raman response.
The parameters we employ for the Hubbard model are typical for the copper-oxygen planes: the (next-)nearest-neighbor transfer integral $t\sim 0.3{\rm~eV}$  ($t'$ $=$-0.2$t$) and the on site Coulomb repulsion $U$$=$$8t$. 
The CDMFT is implemented on a 2$\times$2 cluster and it is solved with a finite-temperature extension of the exact diagonalization method \cite{Liebsch2012}. Previous CDMFT studies have reproduced various essential features of the cuprate phase diagram \cite{Maier2005,Kotliar2006,Tremblay2006,Haule2007,Gull2010,Sordi2012,Alloul2014519}, including the Mott insulator, antiferromagnetism, pseudogap \cite{Huscroft2001,Civelli2005,Maier2005,Kyung2006,Gull2010,Ferrero2010,Sordi2010,Sordi2012a,Sakai2009,Sakai2010,Liebsch2009} and $d$-wave superconductivity \cite{Maier2000,Lichtenstein2000,Maier2005a,Kancharla2008,Civelli2009,Sordi2012,Gull2013}.
However, the optimal doping in the 2$\times$2 CDMFT for which \Tc is maximal ($x \simeq 0.08-0.10$) is smaller than the one ($p \simeq 0.16$) in
experiments. For this reason, we use $x\simeq 0.065$ in the present CDMFT study to discuss the properties of the slightly underdoped cuprate. A quantitative comparison with experiments is therefore not possible and we restrict ourselves to a qualitative one. 

The CDMFT \BAN Raman response displayed in Fig.~\ref{fig:1}(b) reproduces the key features 
found in the experiment. 
First, the  CDMFT results portray well the experimentally observed PP-dip structure in the superconducting state (and they are in agreement with previous calculation with a similar method \cite{Gull2013b}). 
The PP and the dip are respectively located at $\omega\simeq 0.45t$ and $0.7t$.
Secondly, this structure is clearly associated with superconductivity: It diminishes with increasing temperature until 
disappearing at \Tc, as seen in Fig.~\ref{fig:2}(b).
Thirdly, as it can be seen in Fig.~\ref{fig:2}(d), the calculated PP position 2$\Delta$ is almost constant with temperature up to \Tc, 
consistently with the experimental data displayed in the same figure, signaling an unconventional
superconductivity (clearly departing from BCS). Furthermore, the ratio $2\Delta/k_B\Tc\sim 10$ in the CDMFT is rather high as compared to the BCS prediction ($\sim 4.28$), like the experimental value ($\sim 13$). 

\begin{figure}[tp!]
\begin{center}
\includegraphics[width=7.0cm]{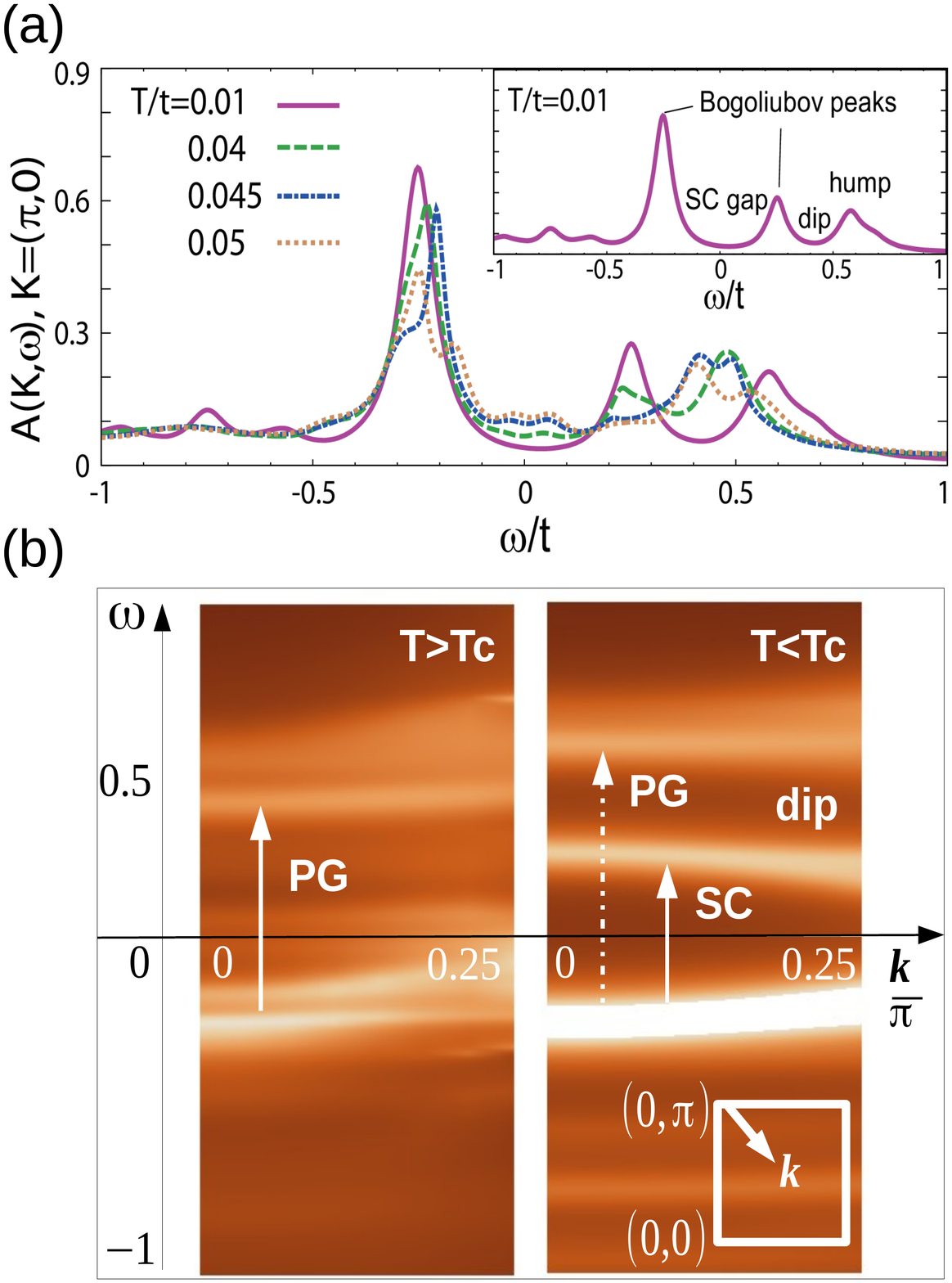}
\caption{(Color online).(a) Spectral function $A(\mathbf{K}, \omega)= -\text{Im}G(\mathbf{K}, \omega)/\pi$ at $\mathbf{K}= (0,\pi)$ for different temperatures below and above \Tc$\approx 0.045t$. 
 (b) Intensity plot of $A(\mathbf{k}, \omega)$ at $T=0.05t$ (left panel) and $T=0.01t$ (right panel) along the $(0,\pi)\to (\pi/4, 3\pi/4)$ cut in momentum space.}
\label{fig:3}
\end{center}
\end{figure}

As Eq.~(\ref{eq:raman}) reproduces qualitatively well the experimental results, we can resort to the single-particle quantities to gain insight into the mechanism originating the PP-dip behavior. To this purpose, we display in Fig.~\ref{fig:3}(a) the spectral function 
$A(\mathbf{K}, \omega)= -\text{Im}G(\mathbf{K}, \omega)/\pi$ at 
$\mathbf{K}= (0,\pi)$ for different temperatures above and below \Tc. Here, \Tc$\simeq 0.045t$ is estimated 
from the disappearance of the superconducting order parameter. 
Figure \ref{fig:3}(b) shows an intensity plot of $A(\mathbf{k}, \omega)$ along the $(0,\pi)\to (\pi/4, 3\pi/4)$
cut in momentum space, showing that the spectral structure of $A(\mathbf{K}, \omega)$ displayed in 
Fig.~\ref{fig:3}(a) is well representative of the antinodal region, which is the most relevant  
to $\chi''_{\BAN}$ via the $\gamma_{\BAN}$ Raman vertex (Eq.~\ref{eq:raman}). 

At $T = 0.05t$, the system is in the normal state. Previous CDMFT studies \cite{Maier2005,Kyung2006}
have established that a pseudogap appears at small doping, as evidenced in Fig. \ref{fig:3}(a)
(yellow-dotted curve) and \ref{fig:3}(b)
by a wide depression around $\omega=0$. The pseudogap edges are located at $\omega=-0.25t$, marked by a well defined 
peak, and at $\omega= +0.4$t, where a wide incoherent hump is observed. 
Since the hump is located inside the Mott gap extending to high energies
($\simeq +6t$ not visible in the figure, see e.g., Fig 11(a) in Ref. \cite{Sakai2010}) we shall call the hump in-gap states. The presence of the in-gap states is a direct consequence of carrier-doping a Mott insulator
without requiring any spontaneous symmetry-breaking \cite{Eskes1991}.
The resulting \BAN Raman response at $T=0.05t$ in Fig.~\ref{fig:1}(b) shows a large incoherent background signal. 
We have shown in a previous work on the normal state that when the pseudogap depression around $\omega=0$ fills in by increasing temperature, $\chi''_{\BAN}$ recovers spectral weight on a wide range at low energy \cite{Sakai2013,Benhabib15}. This is consistent with the experimental Raman response Fig.~1(a) above \Tc=122 K, where the low energy spectral weight is partially restored as the temperature rises up to 180 K. 

Below $\Tc\simeq0.045t$ the superconductivity develops by opening a superconducting gap symmetrically around the Fermi level $\omega=0$ (Fig. \ref{fig:3}(b)). In the conventional BCS mechanism, the spectral weight removed around $\omega=0$ is accumulated at the gap edges where coherent (narrower and with higher intensity) 
Bogoliubov peaks are formed. In preformed Cooper pair scenario the gap already exists above \Tc, and one should just observe the Bogoliubov peaks arising at the pseudogap edges. 
The interesting unusual property is how this is taking place in our system, where a pseudogap-spectral-weight 
depression already exists around the Fermi level above \Tc and it is not particle-hole symmetric \cite{Sakai2013} like the superconducting gap. At negative energy, the lower Bogoliubov peak arises almost at the pseudogap edge, 
in line with the preformed-pair description and as supported for instance in tunneling and ARPES experiments on Bi-2212 
materials \cite{Renner1998,Norman1998,Chatterjee2011,Hashimoto2014}. 
At positive energy however the upper Bogoliubov peak develops at $\omega\simeq 0.25t$, significantly lower than the pseudogap edge at $\omega\approx 0.4t$. 
This process reassures the transfer of spectral weight from low energy ($\omega \simeq 0$) like in BCS,
but  also from higher energies (in correspondence of the pseudogap upper edge  $\omega\simeq 0.4t$), where the dip forms. With decreasing temperature the upper Bogoliubov peak grows and dip deepens, as evident in Fig.~\ref{fig:3} (a). 

A previous study \cite{Sakai2016} has shown that the competition between pseudogap and the superconducting gap can be explained by nontrivial cancellations 
in pole features of the normal and anomalous self-energies, which makes 
possible for the upper Bogoliubov peak to arise at energies where 
the spectral weight has been suppressed by the pseudogap. 
This result advocates in favor of coexisting between the pseudogap and the superconducting gap below \Tc, 
with the latter appearing  smaller \cite{Gull2015} (see Fig.\ref{fig:3}(b)), when observed in the unoccupied side of spectra as in Raman spectroscopy. 

The peak-dip structure on the positive frequency side of $A(\mathbf{k}, \omega)$, displayed in Fig.~3(a-b), 
produces the PP-dip structure in the calculated \BAN  Raman response in Fig.~1(b). As it can be seen from Figs.~2(a) and 2(b), the 
PP-dip structure  in $\chi''_{\BAN}$ is therefore the direct key fingerprint of an unconventional pairing mechanism involving transfer of spectral weight from high-energy states. 
  
In conclusion, we have studied a key temperature-dependent peak-dip relation in the Raman \BAN response of the 
superconducting state of slightly underdoped Hg-1223 by combining Raman experiments and CDMFT calculation.
We reveal an unconventional pairing mechanism originating from the interplay between the superconducting gap and the
pseudogap in the antinodal region. In order to form the Cooper pairs, spectral weight is transferred 
not only from states close to the Fermi level but also from high-energy states located at the pseudogap upper edge. 
The final scenario conveyed here is unusual within the debate on the relation between unconventional superconductivity and pseudogap: while matching on the negative energy occupied side, they appear competing for the same electrons in the positive energy unoccupied side of the electronic spectra, being at the same time friends and foes \cite{Norman2005}.

We are grateful to A. Georges, M. Imada, I. Paul, A.J. Millis and  A. Auerbach for fruitful discussions. Correspondences and requests for materials should be addressed to A.S. (alain.sacuto@univ-paris-diderot.fr), M.C. (civelli@u-psud.fr) and  S.S  (shiro.sakai@riken.jp). 
S.S. is supported by JSPS KAKENHI Grant No.~26800179; B.L. is supported by the DIM OxyMORE, Ile de France.

\newpage

\section{SUPPLEMENTAL MATERIAL}

\subsection*{Peak-Dip Structure In Single Layer HgBa$_2$Cu$_4$O$_6$}

In Fig.1 is displayed the normal (100 K) and the superconducting (12 K) \BAN Raman responses of a slightly UD HgBa$_2$Cu$_4$O$_6$ (Hg-1201) single crystal with a \Tc=92 K  (\Tc (max) =96 K).  The Raman spectra were obtained by using the same laser excitation line (532 nm) than the one used in Fig.1 (a) of the article. 
Contrary to the Hg-1223 material, phonon lines superimposed to the electronic background make more difficult the observation of the pair breaking peak-dip structure. However by subtracting the normal state Raman response from the superconducting one, we are able to detect peak-dip structure (see inset of Fig.1). They are respectively located at $\approx550$ and $\approx900~cm^{-1}$.
This result confirms that the peak-dip structure that we observe in the \BAN Raman response in the superconducting state is not linked to a inter-layer band splitting effect. 


\begin{figure}[ht!]
\begin{center}
\includegraphics[width=9cm,height=6cm]{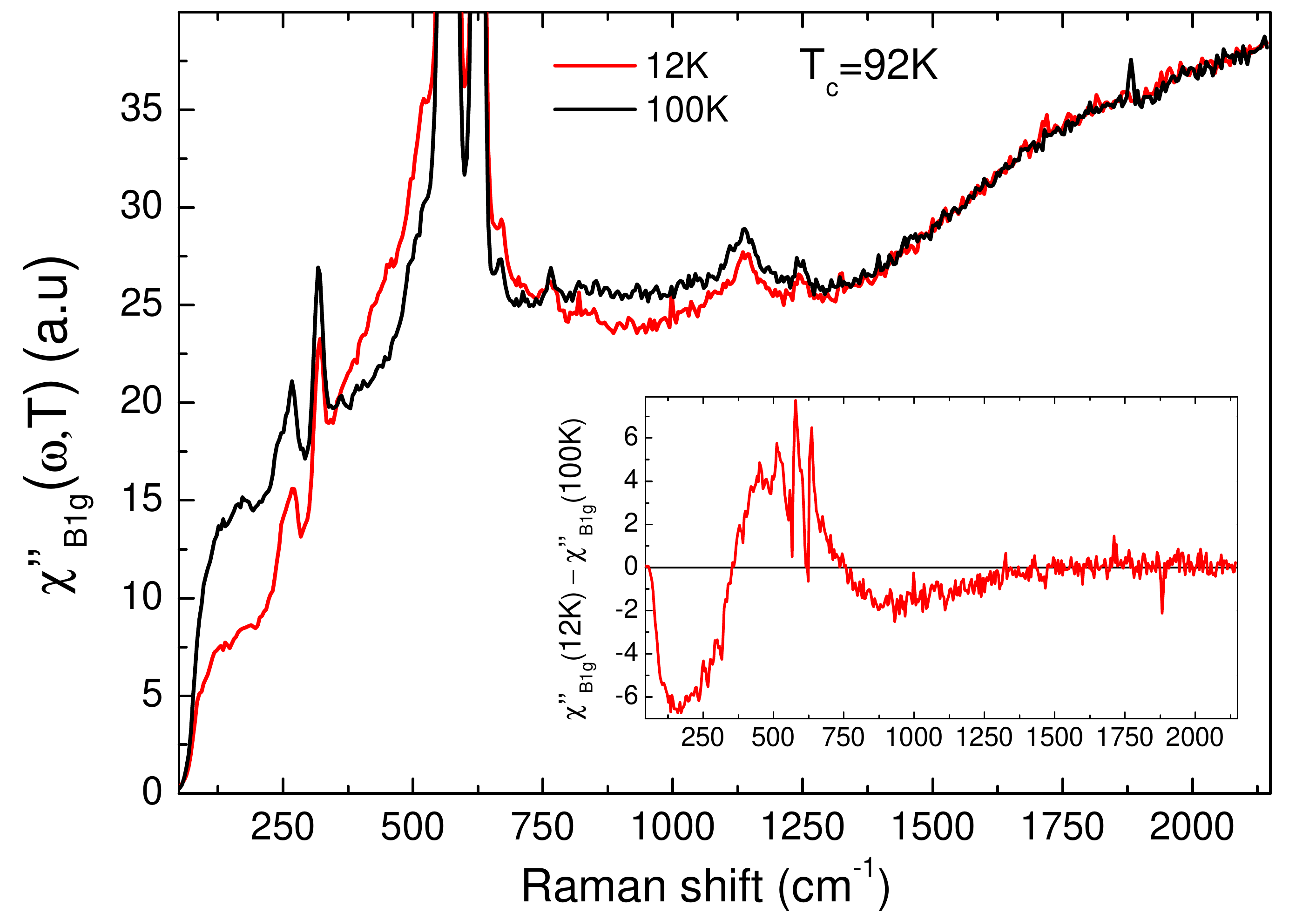}
\end{center}\vspace{-7mm}
\caption{ \BAN Raman response of slightly UD Hg-1201 in the normal and superconducting states. Inset: subtraction between the normal and superconducting Raman responses. The PP and dip are respectively located around 550 ~$cm^{-1}$ and  900 ~$cm^{-1}$. } 
\label{fig1}
\end{figure}

The normal (100 K) and superconducting (12 K) spectra have been measured by using an ARS closed-cycle He cryostat. The laser power at the entrance of cryostat was maintained below $2~mW$ to avoid over heating of the crystal estimated to $3~K/mW$ at $10~K$. The Raman spectra have been corrected for the Bose factor and the instrumental spectral response. They are thus proportional to the imaginary part of the Raman response function, $\chi^{\prime \prime}_{\BAN} (\omega)$.

\clearpage

\bibliography{references}

\end{document}